# Pixelated Interactions: Exploring Pixel Art for Graphical Primitives on a Pin Array Tactile Display

Tigmanshu Bhatnagar

UCL Interaction Centre and Global Disability Innovation Hub, University College London, London, United Kingdom, t.bhatnagar.18@ucl.ac.uk

Vikas Upadhyay

School of Information Technology, IIT Delhi, New Delhi, India, vikas.upadhyay@cse.iitd.ac.in

Anchal Sharma

Department of Design, IIT Delhi, New Delhi, India, anchal.sharma@design.iitd.ac.in

PV Madhusudhan Rao

Department of Design, IIT Delhi, New Delhi, India, pvmrao@design.iitd.ac.in

Mark Miodownik

UCL Mechanical Engineering, University College London, London, United Kingdom, m.miodownik@ucl.ac.uk

Nicolai Marquardt

Microsoft Research, Redmond, Washington, United States and UCL Interaction Centre, University College London, London, United Kingdom, nicmarquardt@microsoft.com

Catherine Holloway

UCL Interaction Centre and Global Disability Innovation Hub, University College London, London, United Kingdom, c.holloway@ucl.ac.uk

Tactile readers with visual impairments read and evaluate the clarity of Pixel Art tactile graphics on a 2D pin-array, Tacilia.

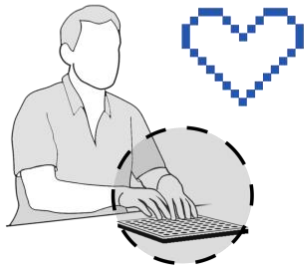

Pixel Art Tactile Graphics on a Pin-array Display are designed and drawn using the following guidelines:

1. Use outlines to create shapes
2. Make single-pixel wide outlines
3. Diagonally connect segments
4. Overlap segments at corners
5. Make proportional segments in curves
6. Make diagonal lines with identical segments
7. Rectilinear element at the apex of curves

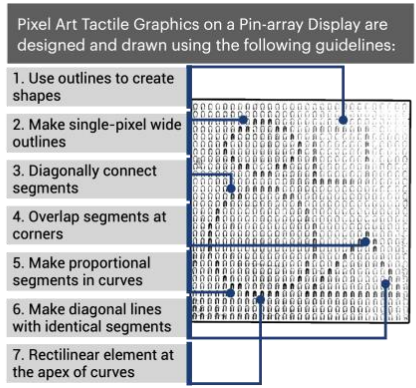

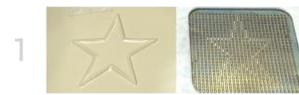

Pixel Art tactile graphics are comprehensible and relatable.

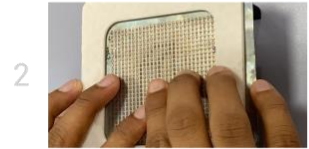

Guidelines provide a framework to design and iterate tactile media on pin-array displays.

Figure 1. Setup of the study and overview of Pixel Art guidelines used to design and draw Pixel Art tactile graphics on Tacilia. Our findings show that basic geometric Pixel Art tactile graphics are comprehensible, and the guidelines provided a framework to design, draw, and iterate tactile media.

Two-dimensional pin array tactile displays enable access to tactile graphics that are important for the education of students with visual impairments. Due to their prohibitive cost, limited access, and limited research within HCI, the rules to design graphical primitives on these low-resolution tactile displays are unclear. In this paper, eight tactile readers with visual impairments qualitatively evaluate the implementation of Pixel Art to create tactile graphical primitives on a pin array display. Every pin of the pin array is assumed to be a pixel on a pixel grid. Our findings suggest that Pixel Art tactile graphics on a pin array are clear and comprehensible to tactile readers, positively confirming its use to design basic tactile shapes and line segments. The guidelines provide a framework to create tactile media which implies that the guidelines can be used to downsize basic shapes for refreshable pin-array displays.

# 1 INTRODUCTION

Teaching practices for students with visual impairments rely on tactile graphics to make images, diagrams, maps, and art accessible. These are simplified translations of visual images that are readable by the tactile sense [50]. For intelligible information, the Braille Authority of North America (BANA) provides guidelines for printed tactile graphics [51], but they do not extend to tactile graphics on refreshable pin array type tactile display. Pin array displays are the refreshable tactile displays on which, tactile information is presented through an array of evenly spaced pins or tactile pixels that selectively pop out above a flat surface to create a tactile bump. Tactile graphics and shapes on these devices are hence created with a series of tactile dots, rather than continuous lines of printed interfaces. Among the many surface haptic technologies, pin array are the closest in terms of their haptic feedback to printed tactile media [30]. There are now a growing number of interactive pin-array displays, and these devices are highly aspirational among students, teachers and professionals who are visually impaired [35].

As technology develops, it is important to have guidelines to ensure repeatability, usability and scalability of information that is presented on the display [52]. Guidelines support an open innovation approach, which has been recommended to scale assistive technology [19]. This is especially important in low-resource environments (LREs) where education resources are severely limited for students with visual impairments, contributing to poor overall educational outcomes [9]. The educational losses compound over a lifetime, restricting the livelihood opportunities of people with visually impairments [15]. Within LREs, digital technologies, such as mobile phones, have bucked the trend of poor technology adoption [22], allowing people with disabilities increased independence and social participation [32]. However, advanced digital interactions, such as pin array displays, have been designed for and with people with visual impairments in high-resource settings. The resulting technologies remain prohibitively expensive for most people with visual impairments globally – a vast majority of whom live in LREs. The exception is Tacilia.

Tacilia is a passive pin-array type display of 729 (27x27) independently addressable taxels made from a single sheet of Nitinol [5]. Each tactile pixel of Tacilia can be manually heated that forces the pixel to bend out of plane to create a tactile effect. Its design has been cocreated with people with visual impairments from LREs. Previous studies have demonstrated that tactile information on Tacilia is comprehensible and tactile graphics can be manually drawn on the display with a hot air jet pencil [5,6]. However, drawings on Tacilia may not always be clear to the tactile sense. Tactile lines may feel broken, jagged, and confusing, making them hard to follow. Hence, to draw tactile graphics on Tacilia and to render pixelated shapes, certain rules and guidelines are necessary so that tactile media can be clearly and consistently presented.

In computer graphics, pixelations of low-resolution images are corrected by anti-aliasing techniques which aim to blend the object's outlines with the background to create the illusion of a smooth line or a curve. The direct application of this technique for pixelated tactile graphics would mean that certain tactile pixels are partially actuated (to a lesser height). We believe this process will not be suitable as the height difference between a full and a partially actuated pixel may not be sufficient to discern. A partially actuated pixel can be perceived as actuated and, therefore, as a part of the shape, causing confusion. Therefore, there is a need to investigate how tactile shapes should be designed for Tacilia, so they are easily understandable. To answer this question, we explored whether Pixel Art – a technique to create detailed graphics on low-resolution visual displays – applies to tactile graphics on a pin-array display. To create Pixel Art, artists follow design guidelines that ensure pixelated outlines appear sharp and curves appear smooth [23,28,38,49]. In computer graphics, algorithms to downsize visual images and automate the pixelation process [17,20,21] such as the Midpoint Line Algorithm developed by Bresenham [8,13] exist. We hypothesize that by considering each pin of a pin-array display to be a *tactile pixel*, the application of Pixel Art on low-resolution pin-array displays can effectively create clear, consistent tactile graphics, given that the pixels are high enough to be clearly perceivable by the tactile sense. In this way, tactile graphics on a pin array display can be considered as a *pixelated alias* of raised line tactile graphics. In the backdrop of this argument, this paper contributes:

- A qualitative evaluation of the application of Pixel Art guidelines on Tacilia, based on a user study that compares the reading experience of pixelated tactile geometric shapes in comparison to a continuous raised line tactile graphics of the same size and
- the qualitative insights about the reading procedure of the Pixel Art tactile shapes and interactions that emerge due to the quick reconfigurability of the tactile display.

We close the paper by discussing the implications of this approach for designing tactile shapes on pin-array displays along with the limitations of this study and scope for future work.

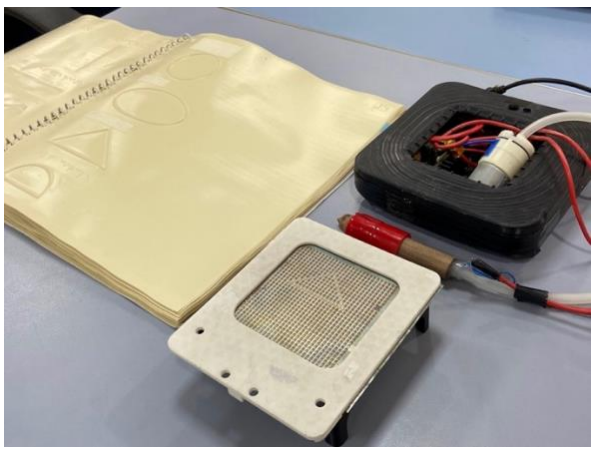

Figure 2. Tacilia, a passive reconfigurable pin-array display (white) with a triangle drawn on it, along with the hot air jet pencil prototype (black + brown pen) and a raised line tactile book (pale yellow) used in the study.

## 2 RELATED WORK

We first describe the challenges when designing in LREs. We then focus on the empirical HCI research related to tactile graphics design on pin-array displays and provide fundamental psychophysical information related to the interpretation of tactile graphics made from a series of dots.

### 2.1 Designing technology in low-resource environments for visually impaired people

Ability-based design [45] is a core design methodology for developing technologies for people with disabilities, exploring what people can do, rather than focusing on their functional limitations. However, what people with disabilities can do is very much dependent on their environment and the resources they have available to them [4]. In LREs access to assistive technology is limited, as are appropriately trained teachers, and infrastructure is challenging. In these environments, advancements in technology have the potential to make a significant impact [39]. However, to succeed, these technologies must be accessible, affordable, and scalable. In this paper, we focus on the latter.

Scalability of solutions is helped by providing guidelines for developing accessible systems [41] e.g. W3C [46]. As was described by Holloway & Barbareschi [53], a strength of the Web Content Accessibility Guidelines (WCAG), is that they provide developers with an actionable framework that is technical and specific. Attention to high-level principles combined with concrete guidelines moves away from the "patchwork of adaptations" that characterized earlier efforts [1]. Such standards have helped scale Braille solutions which allow visually impaired people to read text. For example, the Orbit 20 reader has been developed to conform to the BANA guidelines, can be assembled locally, and was developed with blind people – it has been described as the most inexpensive Braille display and is now in use across Africa [12]. However, this device has limited cell array size and cannot yet handle graphical components. Therefore, a gap remains within tactile graphical interfaces [30], and guidelines for how we render images on low-cost devices will be essential to enable scale of new technologies in this space.

## 2.2 Perception of Tactile Shapes through Pin Array Displays

Existing refreshable tactile displays have an image conversion processor that segments an image, semantically renders it, and maps it to the low resolution tactile pin-array that varies from 60x40 pins [54] to 104x60 pins [55]. This mapping is visually relatable to the original image however, the images may not be clearly comprehensible for the tactile sense [7]. From existing literature, we know that making large graphics [44], eliminating details [36] and presenting only relevant information through the outlines of shapes [3] are some recommendations for creating tactile graphics on pin array type display technologies. However, there is a need to further down sample tactile graphics to represent them intelligible on a pin array display.

Picard and Lebaz [34] report that the ability to read tactile pictures accurately depends on (i) the display modality, (ii) the complexity with which the objects are depicted, (iii) the exploration procedures used and (iv) the availability of semantic information. With Tacilia as the display modality, we provide background on the remaining three attributes that influence the ability to read tactile graphics.

### 2.2.1 Simplifying Complexity

To simplify graphics' complexity, BANA recommends that details should be eliminated and the graphic simplified without losing conceptual information [51]. However, these simplifications are limited to raised line graphics. Further simplification is required to translate raised line graphics to pin array graphics. A straightforward approach would be to increase the size of the tactile graphic to accommodate the finer details as Wijntjes *et al*. [44] demonstrated that larger tactile drawings were recognized more frequently than smaller drawings. However, considering the technological limitations of pin-array displays, increasing the size will significantly add to the cost of making tactile devices. With the present size, Bornschein *et al*. [7] found that downsizing visual graphics to meet the low resolution and size of HyperBraille [35] (10dpi, 60x120 dots) leads to problems in which crucial conceptual details disappear leading to incomplete and confusing shapes. To solve this problem, they discuss the need for a special rendering tool that can re-create an image with respect to the low-resolution of the display whilst maintaining the crucial identifying markers of the image, but do not present any concrete method to do so.

Bellik and Clavel [3] evaluated multiple tactile image rendering methods on the HyperBraille display to identify the fastest and most accurate tactile recognition performance. Experimental analysis from 40 sighted participants revealed that static outlines of shapes, with empty space inside, were the easiest to recognize compared to any other method to present tactile images. Similar results were reported in previous studies with Tacilia [5]. The authors found that blindfolded, sighted participants could accurately determine outlines of basic shapes of a square, circle, and a triangle made on a 5x5 pixel array with a 10dpi resolution. Velazquez and Bazan [42] evaluated the performance of their SMA based tactile display with five blindfolded subjects who had no previous knowledge of braille or tactile graphics. They also found that participants mainly focused on exploring the borders of shapes and faced challenges in recognizing a filled circle. Clearly, existing research provides evidence that tactile graphics on pin-array displays are best read by their outlines.

### 2.2.2 Exploration Procedures and Assimilation of Semantic Information

To identify two discreet tactile dots, the spatial threshold on the fingertip of a reader with visual impairment ranges from 1.0 to 1.5 mm [27]. The inter-pixel distance of pin-array surfaces that are intended to present both braille and tactile graphics will typically follow Braille standards and can range from 2.37 to 2.5 mm [40,56] inter-pixel distance. Therefore, to create the illusion of a raised line on a pin-array, discrete co-located dots must be perceptually connected. Contemporary research has shown that the Gestalt Principles for visual perception [57] are applicable to tactile perception [58]. Discreet but collocated actuated dots separated by gaps can be interpreted as continuous, following the Law of Good Continuation [11], creating the illusion of a line. Chang *et al.* report that tactile elements placed closed together are grouped by our perception and follow the Law of Proximity [10], which would mean that larger gaps between a series of dots separate them into different segments. These laws of perceptual grouping explain the ways in which a series of co-located but discreet tactile pixels of a pin-array display are combined by our perception to form line segments. Moreover, following the line segment over a flat surface provides a temporal haptic input which is eventually consolidated and gets interpreted by the visual processes, as explained by the Image Mediation Model [24,26]. The visualization of the haptic input is correlated to the long-term memory, which also contains the image [31], and hence, it can be recalled. In this way, a pixelated tactile graphic is interpreted.

In summary, perception of tactile graphics on the pin array displays improves when visual images are conceptually simplified, shapes are made by their outlines, and semantic information is available. But other than that, nowhere in the literature could we find any design principles or rules that can facilitate the proper rules to create tactile shapes on pin-array displays so that they are read accurately. We need fundamental and replicable rules that are agnostic to a display to logically actuate only the required dots on a pin-array. Hence, we explore the application of Pixel Art for this purpose using a 2D pin-array display, Tacilia.

## 3    WHY AND HOW TO IMPLMENT PIXEL ART?

In digital graphics, scan-conversion algorithms for graphical primitives compute the coordinates of the pixels that lie on or near an ideal thin line on a 2D rasterization grid of pixels. Considering a 1-pixel thick approximation for a straight line with a given slope, the sequence of pixels representing the line in a set array is decided by the Midpoint Line Algorithm developed by Bresenham [8,13]. The algorithm computes and finds the closest pixel in the array to the ideal line. It does so by using an incremental technique in which, depending upon the slope of the line, the coordinate of x or the y-axis is incremented by one, and the closest pixel to the point of intersection between the two possible pixels on the incremented value is selected. In this way, an approximate line segment on a 2D pixel array is rasterised. In situations when the line segment intersects exactly in the middle of two potential pixels, the endpoints of the line must be adjusted to make a choice for selecting a suitable pixel. This algorithm has also been adapted to raster circles and ellipses in which the drawing process is further improved by producing eight-way symmetric segments. Pixel artists also adopt the algorithm to create intelligible low-resolution art. It is a form of digital art where a high level of detail is achieved at a low image resolution. There are several recommendations to render Pixel Art [38,49] and Keddy [17] and Yu [28] list guidelines for artists, so that jagged lines appear sharp, and curves appear smooth.

The 2D pixel grid of digital graphics can be considered analogous to a tactile pin array. Each pin of the pin-array display is regarded as a pixel. Hence, only the pixels that follow the Pixel Art guideline are actuated to create a shape. In this way, we can recreate any pixelated visual graphic in its tactile form on any pin array display. In the following exploratory study, we create basic and complex geometric shapes for each participant using the Pixel Art guidelines on Tacilia. Table 1 presents an overview of the Pixel Art guidelines, which have been the basis for multiple pixelation algorithms for digital visual graphics [14,17,20,25]. Table 1 also provides the tactile representation of a pixelated line segment on Tacilia based on each guideline.

Table 1. Overview of the Pixel Art Guidelines and their Tactile Graphic Implementation.

| | Pixel Art Guideline | | Tactile Graphic Implementation |
|---|---|---|---|
| 1 | **Single Pixel Wide Outlines** <br> All shapes should be made using just a single pixel wide outline. Lines with multiple thickness should be avoided and all extra pixels should be removed. | 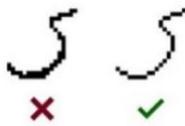 | 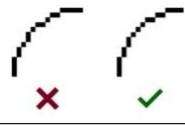 |
| 2 | **Diagonal Connections between Segments** <br> In diagonals lines and curves, pixels of horizontal and vertical line segments should only be connected diagonally. | 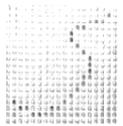 | 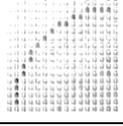 |
| 3 | **Equal Segments in Diagonal Lines** <br> Diagonal lines are made up of a series of smaller lines and all the smaller line segments should be identical in length as much as possible. | 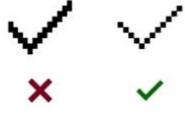 | 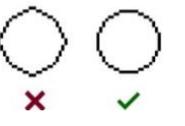 |
| 4 | **Proportional Segments in Curved Lines** <br> For curved lines, the length of the segments should be proportionally reduced towards the apex of the curve. | 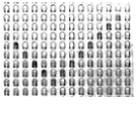 | 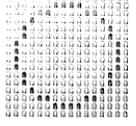 |
| 5 | **Straight Line Segments at the Apex of Closed Curves** <br> In closed curves like a circle, the top, bottom, left and right can be straight and identical in size to avoid any blips. | 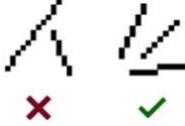 | 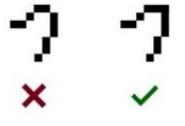 |
| 6 | **Overlapping Pixel at Corners** <br> Outlines on vertices of angular shapes should not be connected only diagonally and must have an extra actuated pixel at the corner to create a sharp point. | 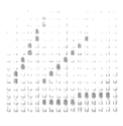 | 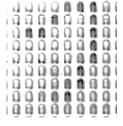 |

## 4  METHOD

We describe the method to evaluate how people with visual impairments read Pixel Art based basic geometric, compound, and complex tactile shapes on Tacilia. The aim of the study is to evaluate the implementation of Pixel

Art to design, draw and read pixelated tactile graphics on a pin array. The Research Ethics Committee from UCL (UCL REC 18925/001) and Institute Ethics Committee from IIT Delhi (IEC P-086) provided the ethical clearance to conduct this study.

## 4.1 Participants

Eight visually impaired adults from India participated in the study (Table 2), of which six were male and two were female (mean age = 28.6 years, SD = 4.4). Six of the eight participants were congenitally blind, one had gradually lost vision up until 14, and one had very limited light perception since birth. Six participants were right-handed, while two had a dominant left-reading hand. Each participant was neurologically healthy and reported no problems with the sensory integrity of their fingertips.

All were recruited through Saksham Trust, which is an organization for the visually impaired in New Delhi. The recruitment was done through purposive sampling, and we included participants who were above the age of 18 and were braille literate. Only adult participants were included in the study due to ethical reasons. In addition, adults may have more experience and familiarity with tactile information compared to children who's limited experience and cognitive abilities may affect the ability to interpret the tactile display. We also included participants with braille literacy to capture their expertise in tactile reading. We anticipated that the experience with braille would make the participants more attuned to the nuances of tactile information, allowing them to provide detailed feedback. They also represent a significant segment of the population for whom the results of this study are most relevant. Children with visual impairments under 18 or adults who had their education solely with audio or low vision educational aids such as large print documents and magnifiers were excluded.

To our surprise, none of the participants had any prior experience with pin-array tactile graphic displays. Three participants had extensive experience reading printed tactile graphics as they were involved in earlier research projects. Two participants had limited experience from their schooling, while three had no experience in reading tactile graphics. Hence, we can confidently say that reading Pixel Art on Tacilia was a novel experience for everyone.

Table 2. Participants for the study.

| Participant | Age | Gender | Visual Impairment | Braille Literacy | Tactile Graphic Experience |
|---|---|---|---|---|---|
| P1 | 22 | Male | Congenital | Yes | No |
| P2 | 27 | Male | Gradual loss of vision | Yes | No |
| P3 | 28 | Female | Congenital | Yes | Yes, extensive |
| P4 | 32 | Female | Congenital | Yes | Yes, extensive |
| P5 | 37 | Male | Congenital | Yes | Yes, limited |
| P6 | 29 | Male | Congenital | Yes | Yes, extensive |
| P7 | 28 | Male | Congenital | Yes | Yes, limited |
| P8 | 26 | Male | Limited light perception | Yes | No |

## 4.2 Apparatus

Tacilia is presented in Figure 3a and is compared to a raised line tactile graphic book (Figure 3b) acquired from Raised Line Foundation [59]. The tactile book was made by thermoforming continuous outlines of shapes on a thin thermoplastic sheet, which is one of the conventional techniques for creating raised-line graphics. The book is currently in use at special and inclusive schools for children with visual impairments in India. The height of the raised line on the thermoplastic sheet was 1mm with a width of 2mm in comparison to Tacilia where the height of each tactile pixel is 0.4mm and its width is 1.2mm.

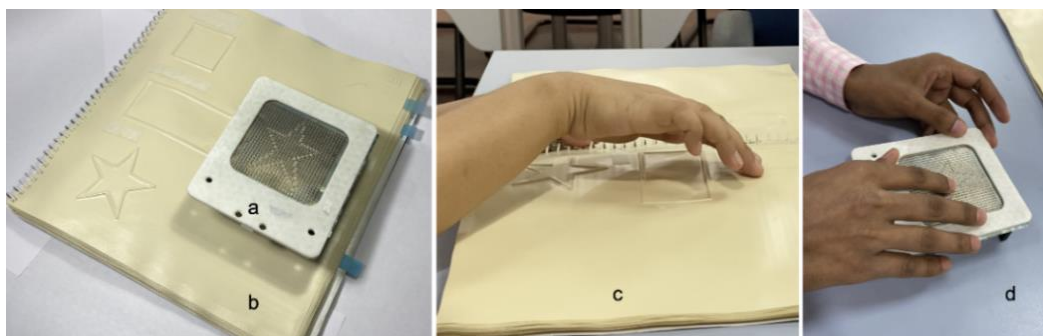

Figure 3. (a) Tacilia (b) Raised line tactile book, (c) reading on the tactile book and (d) reading on Tacilia.

We anticipated that prior experience of reading shapes on printed tactile media might lead to a better reading performance on the book compared to the pin-array display, which was being experienced for the first time. Furthermore, bolder, and continuous raised lines would be clearer than thin and light-pixelated graphics. While we acknowledge the differences, it is important to clarify that the study did not focus on evaluating this specific pin-array display device in comparison to a thermoformed tactile book. Instead, the aim of our study is to evaluate the effectiveness of using Pixel Art to design and read tactile shapes agnostic to a pin-array display. From this point of view, Tacilia was used as the reconfigurable pin-array display because Pixel Art tactile graphics can be created and evaluated on it, and a tactile book was used for within-subject comparisons to evaluate the reading performance.

### 4.3 Procedure

Participants were invited to a well-ventilated room and seated in front of a table with the test set up where two researchers facilitated the experiment. Informed consent was taken by the participants to conduct the study and record their interactions. We conducted a three-part study to qualitatively evaluate the implementation of Pixel Art on a pin array display. No additional time was given to train on Tacilia as we wanted to observe their out-of-the-box experiences with Pixel Art tactile graphics.

#### 4.3.1 Part One

First, we wanted to evaluate the comprehensibility of Pixel Art tactile shapes on the pin array display in comparison to the raised line tactile drawings on the book. Six basic geometric shapes were selected to evaluate the clarity of Pixel Art tactile graphics compared to the same shapes of the same size on the raised line tactile graphic book (Table 3). These basic shapes were used as a proxy to implement Pixel Art in tactile graphics as line segments in them used a combination of Guidelines 1 to 6.

In this part of the study, no participant was given any hint or information about the shape or the medium they were about to see. Participants we asked to recognize the basic shapes on the two tactile mediums. We drew one shape at a time on Tacilia and intermittently showed a shape on the tactile book. Shapes on the two mediums were presented according to a balanced Latin square stimuli randomization to avoid order effects between the book and the display. For both the book and the display, participants were requested to read the shape only tactually but were given freedom to explore it by the pattern and procedure of their choice. Everyone was asked to say 'start' when they commenced reading, to think-aloud about their reading experience while reading, and after reading, utter the name of the shape to indicate that they have finished the reading. After reading, participants were asked to elaborate on the clarity of the shape and highlight any problems that they encountered in identifying the shape.

Table 3. Shapes chosen for comparison on the two tactile mediums.

| Shapes/ Medium | Square | Rectangle | Circle | Ellipse | Triangle | Star |
|---|---|---|---|---|---|---|
| Tactile Book | 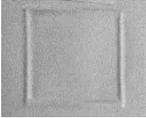 | 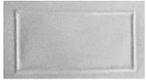 | 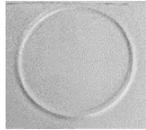 | 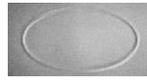 | 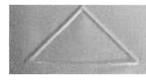 | 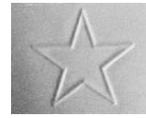 |
| Size (bounding rectangle (lxb)) | 25 x 25mm | 25 x 50mm | 35 x 35mm | 20 x 50mm | 30 x 60 mm | 50 x 50mm |
| Pin-Array Display | 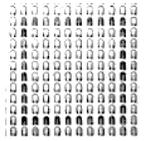 | 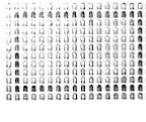 | 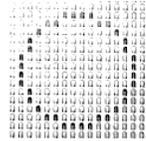 | 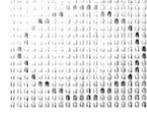 | 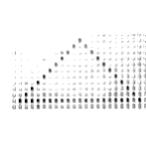 | 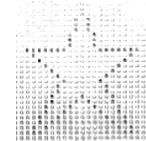 |
| Size (bounding rectangle (lxb)) | 25 x 25mm | 25 x 50mm | 35 x 35mm | 35 x 47.5mm | 27.5 x 55mm | 62.5 x 47.5mm |
| Guidelines | 1 & 6 | 1 & 6 | 1, 4 & 5 | 1, 4, 5 | 1, 2, 3 & 6 | 1, 2, 3 & 6 |

### 4.3.2   Part Two

In the second part of the study aimed to gather more data about the reading procedure on the tactile display in addition to the previous six shapes. Therefore, participants were then given two additional compound shapes (Table 4) to read and identify only on Tacilia. Beyond the comparison of basic shapes, these two additional Pixel Art graphics have a combination of line segment types and in multiple orientations that further expands the application of Pixel Art to design tactile shapes. Participants were again instructed to recognize the shape only tactually and asked to indicate when they start and when they finish for us to record the time. During the time of the exploration, participants were instructed to think aloud what they feel and what they anticipate the shape to be. No information about the shape was given to the participant before reading, but they were asked to share any prior experience with the shape after they had finished reading the shape.

Table 4. Additional compound shapes.

| | Pentagon | Heart |
|---|---|---|
| Compound Shapes |  |  |
| Size (bounding rectangle (lxb)) | 42.5 x 42.5mm | 37.5 x 37.5mm |
| Guidelines | 1, 2, 3 & 6 | 1, 2, 3, 4, 5 & 6 |

### 4.3.3 Part Three

In addition to these predefined geometric shapes, we asked each participant if they would like to challenge us with a tactile graphic they would like to explore on the display, which we drew for them on-the-spot. This part of the study aimed to qualitatively reflect on the effectiveness of the guidelines to design arbitrary tactile shapes manually. We were also keen to observe the emergent interactions that this new interdependent interaction would provide. Hence, participants five of the eight participants instructed us to draw a tactile graphic, while three dropped out from this part of the study due to time constraints. The shapes that were designed and drawn on the spot are presented in Table 5. Their sizes were similar to the previous shapes and were designed on-the-spot using Pixel Art.

All the shapes (6+2+5) in the three parts, irrespective of their complexity, followed Guideline 1 and were made from only a single pixel-wide outline. In addition, the diagonal lines of a triangle, pentagon, heart, star, cuboid followed Guidelines 2, 3 and 6 in which all their diagonal line segments were made of smaller but identical segments that are connected diagonally, and the segments only overlap on a single pixel at the vertex. Guideline 6 is also used for the rectilinear shapes of a square, rectangle, and the axes of the sine curve. Curved line segments of a circle, ellipse, heart, flower, smiley face, and the sine curve follow Guidelines 4 and 5 in addition to Guideline 1.

Table 5. Graphics designed on-the-spot.

| | Sine Curve | Smiley Face | Flower | Cuboid | English Characters |
|---|---|---|---|---|---|
| Tactile Graphic Challenge |  |  |  |  |  |
| Size (bounding rectangle (lxb)) | 67.5 x 67.5mm | 37.5 x 37.5mm | 57.5 x 50mm | 40 x 32.5mm | 50 x 65mm |
| Guidelines | 1,2,3,4,5 & 6 | 1, 4 & 5 | 1, 4 & 5 | 1,2,3 & 6 | 1,2,3,4,5 & 6 |

### 4.4 Analysis

In total, we had 13 hours of video recordings that were used to inform qualitative insights. For the first part of the study, from the videos, we calculated the time from the start till the time the participants either said the name of the shape aloud or gave up and were unable to recall the name of the shape. We noted the time it took to read the six shapes on both the mediums by the eight participants. In total, we had 96 readings of time for analysis. We compared the time it took for each shape on the two tactile displays and measured the accuracy of their recognition. We also analyzed the explanations and subjective feedback that the participants provided. These verbatim were crucial for those scenarios in which the participant was unable to recognize the shape or corelated it with other shapes that were experienced in real life. For the second part of the study, in which we included two more compound shapes on Tacilia, we inductively coded the eight shapes that were presented to each participant. Therefore, we had 64 samples of tactile interactions with Pixel Art tactile graphics. We observed the procedure each participant followed to acquire the tactile information. We also evaluate the perceived clarity of the pixelated line segment types (rectilinear, curves and angular). For the third part, the researchers made notes on the drawing procedure to create arbitrary shapes, which were being evaluated by the participants and observed the emergent interactions due to this interdependent activity. These qualitative notes and observations were later discussed among the researchers. The findings from this analysis are discussed ahead.

## 5  FINDINGS

The evaluation of the implementation of Pixel Art on Tacilia are described in four sections. The first section presents the quantitative and qualitative data that comparing the reading patterns and reading time on the two tactile modalities. We report slower reading speed for pixelated tactile graphics, but similar reading patterns on the two mediums. The second section reports the reading procedure for pixelated tactile shapes; that has four sequential steps. The third section provides qualitative and qualitative evidence about the clarity of Pixel Art based line segments. The fourth section reports reflections about designing Pixel Art shapes manually and observations about the emergent interactions due to the interdependent approach.

### 5.1  Reading Patterns are the same, but Reading Time varied with the interface

Firstly, we observed no significant difference in reading patterns [47] within-subjects between the pixelated lines and the continuous raised lines. On both tactile mediums, P2, P4 and P7 had one finger fixed over a line segment or a vertex of the shape, and only the other hand's index finger moved over other line segments to acquire information about the shape. P1 and P5 switched fingers on static points while the other finger moved over the tactile path on both the mediums and across all the shapes. P6 and P8 went back and forth over multiple segments simultaneously with two fingers of both hands. P8 also used pinching actions to estimate the proportions of a shape. P3 used a unique single-hand multi-finger reading pattern to get a quick overview which also had pinching actions across line segments to understand proportions. Participants did not adapt or invent a new reading pattern to read Pixel Art tactile graphics. P3, for instance, being an experienced tactile graphic reader, compared the clarity of the two mediums and found pixelated graphics to be different but equally clear:

P3 (after reading a rectangle on Tacilia and comparing it to the book) - "This is a rectangle. It is very clear. Anyone can make it out. The only difference that the book has a straight line, and this has dots. Otherwise, it is very clear."

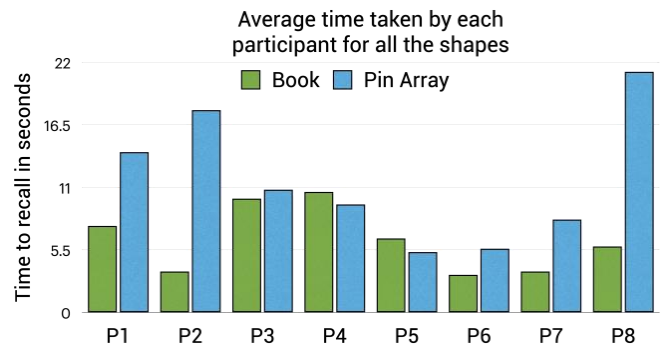

Figure 4. Average time taken to identify all the shapes by participant, note the difference between the times taken by P3 – P7 on both the mediums and the times taken by P1, P2 and P8.

On the first interaction with Tacilia, participants noted that the lines appeared 'light' and 'thin'. The 'lighter' lines could be attributed to the lower actuated height (0.4mm) and the inter-pixel distance in Tacilia compared to the continuous raised line of the book. Participants said that assimilating information on the display required greater focus. Participants with prior exposure to tactile graphics performed better with Pixel Art in terms of the recall times and accuracies compared to participants who had little or no experience to tactile media in general. P3, P4, P5, P6 and P7 had prior experience in reading tactile graphics (Figure 4). Their reading times on the pixelated display are less on average than P1, P2 and P8 who had limited experience with tactile graphics before. However, P1, P2 and P8 have comparable reading times on the book to other participants. This observation suggests that thin, light, and pixelated lines posed difficulties in reading shapes on the pixelated display.

For example, we observed that P1's and P2's finger movements were slower on the pin array that led to their higher reading times. Most participants and even P1 and P2 who had no prior experience with tactile graphics could adjust to the *lightness* of the interface and were able to accurately interpret or corelate all the six basic Pixel Art shapes. For example. P7 saw the star for the first time on the tactile book and then, to his astonishment, was able to correlate the shape on Tacilia demonstrating a comparable depiction of the shape through a series of dots:

P7 (while reading the Pixel Art star) – "This is the same star as I saw in the book. Yes, this is a star. This is the same figure as I was shown in the book."

### 5.2  Reading Pixelated Tactile Graphics has Four Steps

Existing literature has reported that sighted readers acquire information in a process flow called Whole-to-Part learning. In contrast, tactile reading happens in the opposite direction; readers instead get information from parts of the tactile graphic that they touch and assimilate information from these parts in a sequence to understand the whole picture [48]. Budling on this phenomenon, we present the sequential reading procedure that is discussed ahead in detail (Figure 5).

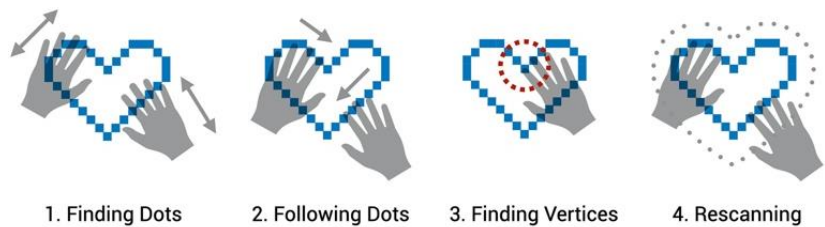

1. Finding Dots    2. Following Dots    3. Finding Vertices    4. Rescanning

Figure 5. Reading procedure for Pixel Art tactile graphics

### 5.2.1   Step 1 – Finding Tactile Pixels

In the first interaction with the Pixel Art tactile shape, participants moved their fingers everywhere on the surface with vertical, horizontal, and diagonal reciprocating actions to find the actuated tactile pixels. Participants distinguished between the actuated tactile pixels and the underlying flat surface of Tacilia and then intuitively followed the adjacent actuated pixels to assimilate the tactile information. Two participants who had no prior experience with tactile graphics initially regarded the pixelated graphic as braille dots and started reading braille characters forming from a few unintended combinations. This demonstrates that novice tactile readers, may find it challenging to discern braille from the tactile graphics. Therefore, rules to juxtapose graphics with labels is necessary for clear description of the shapes because after that, readers were able to differentiate pixelated graphics from braille, which is highlighted in the following moment:

> P1 (first interaction with Tacilia) - "What is this, do we have something here? Ah we must read all these dots, okay. Have you written something here in braille? (Experimenter responds) Ah so, we must recognize this shape these dots make. Okay, I will try."

### 5.2.2   Step 2 – Following Tactile Pixels

The single pixel wide line (Guideline 1) was effective in guiding the fingers to follow and connect a series of tactile pixels. The contrast between the actuated pixels and the adjacent flat surface was felt on either side of the fingertip made the lines clean and distinct where actuated tactile pixels in a predictable continuation created a suggestion of line segment that guided the fingers to follow it. However, as mentioned before, all the participants remarked that lines made by tactile pixels were 'light' as captured in the following comment from P2 who was exploring tactile graphics in both mediums for the first time:

> P2 (first interaction with Tacilia) - "Which shape, am I touching at the right place? Yes, there is something made in braille or in dotted lines. (After following the lines) From some places, I think this is a star, but I will have to see it properly because the dots are not that high. If the dots were higher then I will be able to identify it easily."

We observed that in shapes such as the pentagon and heart, which had multiple types of diagonal lines and the heart that had two curved line segments and two diagonal segments, no participant had any difficulties in identifying the changing nature of line segments (Guideline 1, 2, 3 and 6). For example, P4 and P8 could detect the different diagonals and sides of the pentagon, though they were not able to recall the name of the shape.

P4 (while reading the Pixel Art pentagon) - "Yes, the lines are clear, that is why I can count the number of sides. The corners are quite sharp, so I can understand it, but what was this called?"

### 5.2.3    Step 3 – Vertex Identification

Vertices were found to be crucial markers and anchor points for the fingers. Each vertex in Pixel Art graphic had one pixel in common between two lines according to Guideline 6. The overlap made the *change* in segments sharp and apparent. Participants confirmed the location of a vertex by slightly moving their fingers colinear to a line segment but ahead of the corner. Finding no new tactile pixels perhaps created a confident assumption that the segment terminates at the location and new segment in another direction has started. Participants were even able to count these vertices to determine the geometric shape. For example, P2, P3 and P4 counted the vertices which were scanned in Step 1 and were able to accurately guess the rectangle, triangle, and the pentagon, skipping Step 2.

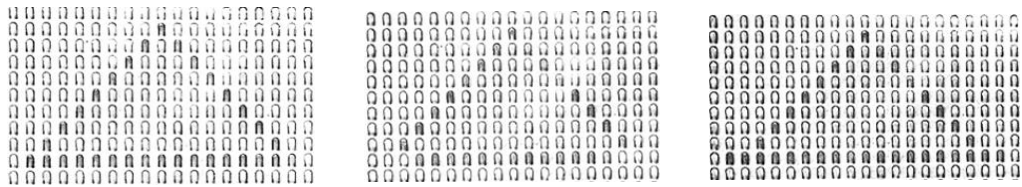
Figure 6: Three versions of tactile triangles.

We explored the effect of changing the design of the vertex by extending and reducing the length of the horizontal base of a triangle (Figure 6). We increased and erased one pixel at the two bottom vertices. In both the scenarios, participants were able to detect the changes to the shape. They perceived the triangle as distorted when the pixels were erased and thought that it may be a different shape, while the extended line was regarded as an error. This exploration indicated that only a single pixel overlap (Guideline 6) is required and is sufficient to create a sharp corner for pixelated tactile graphics.

### 5.2.4    Step 4 – Rescanning:

On both the tactile mediums, we observed that most participants rescanned the entire shape just before, while and even after verbalizing their response. For P1, P2, P5 and P8 it was a common procedure for every shape on both mediums. Some participants employed a different reading pattern at the rescanning stage, for instance, P1, after identifying and verbalizing the pixelated circle after following the curved line, started using multi-finger pinching movements to rescan and confirm the roundness of the shape.

P1 (while reading the pixelated circle) – "This is a circle, like a wheel. A little bit, like…. umm (using pinching action) Yes this is a perfect circle!"

Upon further inquiry, P2 and P8 shared that rescanning was necessary to gain confirmation about the presented graphic. However, after these four procedural steps, participants were able to perceive the shape, even if they had challenges in recalling its name.

### 5.3 Clarity of Pixel Art Line Segments

Most elements of a visual image are translated into tactile graphics through an embossed line, making raised lines the most important element for printed tactile graphics [2]. We now dive into the analysis for each type of line segment presented to the participants.

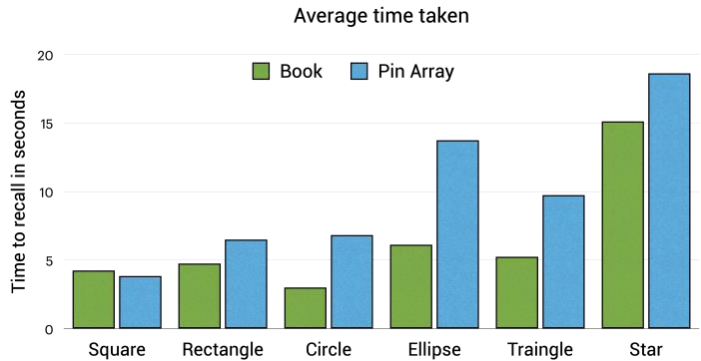

Figure 7. Average time taken to identify each shape. Not the higher times to read ellipse and the star.

#### 5.3.1 Rectilinear Pixel Art Segments

On Tacilia, straight, single-pixel wide (Guideline 1) horizontal and vertical line segments and shapes made using them were clearly identifiable with high confidence. Although the time to read shapes was higher on Tacilia in comparison to the book, no participant had difficulty in identifying rectilinear shapes on Tacilia and the accuracy of recall of a square and rectangle was 100%. The single pixel overlaps were regarded as a clear vertex (Guideline 6). Furthermore, no challenges were encountered in identifying the pixelated horizontal or vertical line segments in other shapes (triangle, pentagon, sine curve, cube and the letters). Here, we would like to mention for a participant, even the rudimentary rectilinear shapes were identified by their descriptive memory. P1 had never seen tactile graphics in school but remembered the description of shapes taught to them. Hence, after following the pixelated segments of a square on Tacilia, P1 was able to correlate the tactile sensation to the imagination that was created by the memory of the shape's verbal description [37]. This is an important observation because it demonstrates the clarity with which the pixelated shape was comprehended by a participant who had no experience with tactile graphics.

> P1 (while reading the Pixel Art square) - "This is a square. Oh! I have never seen it in a tactile form like this. I have studied in a village school, so I only remember its description that if two lines are parallel and each line is of the same length connected at ninety degrees then it's a square. So, it's exactly like that."

#### 5.3.2 Diagonal Pixel Art Segments

Diagonal connections between smaller identical segments created a clear tactile percept of a diagonal line (Guideline 2, 3). The time taken to recall shapes with diagonal segments on Tacilia was again higher than on the tactile book. However, the accuracy of determining triangles on both the mediums was 100% while a pixelated star was accurately identified by seven of the eight participants. The experience of identifying sharp vertices and connected lines surprised some participants, for example:

P4 (while reading the pixelated star) - "This is a star, oh! You have managed to come quite close to it. I was not sure if these bends would be managed, I was not sure somewhat. But this is quiet there. If someone has seen this on paper and clearly remembers it, then I am sure they will be able to see this."

Most participants had no difficulties reading the diagonal line segments of a triangle, pentagon, and the heart. However, the larger gaps between diagonally placed adjacent pixels compared to adjacent pixels in rectilinear lines was noticeable to P6, to which the participant mentioned that the '*intensity*' of dots in diagonal segments was not the same as in the horizontal or vertical line segments. In a pixeled array, a diagonal line will be sparser compared to horizontal or vertical lines due to the fixed inter-pixel distance. Still, it was not problematic for any other participant and despite the changes in perceived intensity, P6 was also able to accurately identify all the shapes with diagonal segments. For example, P6 had earlier seen the 2D tactile representation of a cuboid in its raised line form and asked us to draw the same. With the tactile memory of seeing a cuboid before, P6 was able to identify the three surfaces of the 2D representation of a cuboid on Tacilia and remarked the following:

P6 (while reading the Pixel Art cuboid) – "Yes you are showing me a 3D thing. This is a surface at the top, this one is on the side and this one is on the front. I can see three surfaces, but the bottom surface is not there and the one at the back is not there. I think that will not even be possible to show. I have previously seen this shape in printed tactile form, that is why I am able to tell you this."

### 5.3.3   Curved Pixel Art Segments

The proportional shortening segments in a curving line (Guideline 4) created a smooth turning effect which participants enjoyed reading. For six out of the eight the interpretation of the Pixel Art circle and ellipse was clear. The average time to recall the circle on the tactile book was again less than on Tacilia (Figure 7). The ellipse, in particular, took longer reading time on average on Tacilia compared to the tactile book (Figure 7). The longer time to identify ellipses on both mediums was mainly due to the general unfamiliarity with the shape's name, even if it was perceivable. This phenomenon has been previously studied in [16] and is illustrated in the following quote:

P6 (after reading the pixelated ellipse) – "This is rounded; this is cylindrical. Although it is round, it is quite long and because of the long length, we can only call this a cylindrical shape. But what did you want to make? (Experimenter answers) Okay ellipse yes, a flattened circle yes that is more exact. I was able to clearly read the shape and yes this is elliptical. I just said cylinder because in mind I did not know that this is called elliptical."

For some participants who did not know the name of the curved shape, the pixelated tactile information was correlated to the contours of other physical objects they have encountered in life. This is captured in the following comment:

P5 (after reading the pixelated ellipse) – "This is a circle, but it does not look like it. I feel that there is no side protruding. What is this… like an egg. Yes, I can follow the shape, but I have never seen a shape like this in 2D nor know its name."

P2 asked to see a mathematical graph that he had seen as a school student when he still had sight, and we presented him with the basic sine curve function. P2 had never seen a graph in its tactile form, but without any hints or guidance, they could explore the dotted line segments of the graphic and its axes and interpret the positively inclined curvature of the function, the origin, and the cartesian quadrants.

P2 (while reading the Pixel Art sine curve) – "This line starts at the negative x-axis goes into the area of negative y-axis and then goes up and crosses the origin into positive x-axis and the turns down. The x and y axis are very clear, and I can follow the curve. It is okay!"

However, in contrast, a few participants faced challenges in reading curves on the pixelated interface, that is captured in the following reflection by P8, who took the most time to read on Tacilia:

P8 (after reading the pixelated pentagon) – "I feel that if there is a rectangle, a triangle, square or even a pentagon, there is no difficulty in understanding the shape. The straight and diagonal lines are clear. I think I have some difficulties in circle and ellipse, especially after the line turns."

In general, making curves and diagonal lines on low-resolution pixel array is challenging [13]. We noticed that the apex segments of curves made using Guideline 5 confused two participants by making them think that the shape has straight lines (Figure 8). P3 and P8 second-guessed a circle to be an octagon and a hexagon, respectively but were underconfident. Upon further enquiry, P3 expressed that an octagon was also not fully apparent because of the curvy lines between the straight lines, and that is why the participant had guessed it to be a circle in the first place. However, this limitation of Pixel Art curves was learned by all the participants. The two participants who had difficulties in reading the first curved shape had no difficulty in identifying the second curved shape, which P3 explains:

P3 (after reading the pixelated ellipse) - "This is an oval shape. Yeah, but you must guess because here (pointing at top apex) and here (pointing at the bottom apex) there are lines. But I think if someone uses this frequently, then it can be learnt easily. Like I have not used it much still, I am able to guess it. If I can guess it, then other people can also guess it. Then even the circle can be read easily."

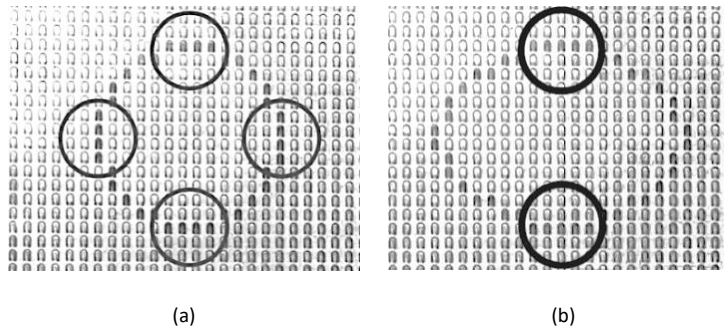

(a)                                          (b)

Figure 8. The straight-line segments in a circle (a) and in an ellipse (b) caused confusion initially.

## 5.4 Emergent Interactions facilitate Interdependence and Learning

The combination of pixelated graphics and the ability to iterate them quickly led to notable interactive experiences. These interactions resonated with the paradigm introduced by Bennett et al. [4] who emphasize that an individual's relationship with the environment is mediated by ATs and relationships with people who collectively work to create access. The refreshability of Tacilia coupled with the ease of implementation of Pixel Art guidelines to create arbitrary shapes enabled an iterative creation of tactile shapes and learning through the experiences. These two emergent interactions are described ahead.

### 5.4.1  Iterative Co-Design of Tactile Graphics

Bornschein et al. [7] have argued that involvement of users of tactile graphics in the early stages of tactile graphic production will increase the reliability of the tactile graphic and speed up the overall production time. Therefore, for early evaluation of graphics, refreshable tactile graphics displays are necessary. We reflect on a scenario in which P3 asked us to draw and present a smiley face icon she has used in text messages.

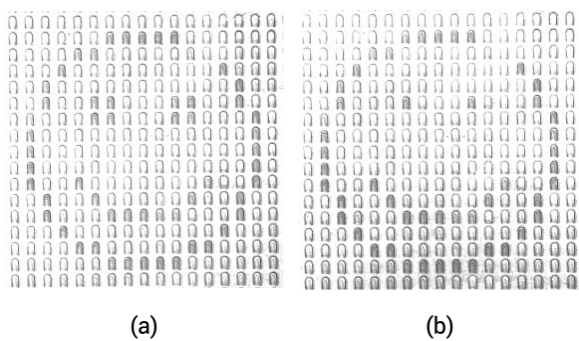

(a)                              (b)

Figure 9. (a) Initial smiley face, (b) Iterated smiley face

We drew a smiley face based on our mental image, using Guideline 1, 4 and 5. Drawing a circle was straightforward, and then inside the circle, two eyes were created that depicted by four co-located pixels to show big apparent eyes (Figure 9a) and a smile was designed with a semicircular arc. After reading through the segments of the face one by one, P3 was able to interpret the round contour of the icon, the semicircular smile but not the eyes. P4 remarked that four dots made a square, and as eyes are round, they were not clear. She suggested that a single dot would be a more accurate representation. We erased the eyes and made them from a single dot (Figure 9b), which P3 enjoyed exploring, as captured in the following remark:

P4 - "This has four dots here, four dots here and then this semicircle and the circle. A smiley face! But here for an eye, there are two squares instead of a dot. If you just make the dot instead of these four dots, then it will be easy. (After iterating and presenting the new smiley face). Wow okay yes, this can be seen easily."

Similarly, P5 wished to see a flower on Tacilia. So, we drew four ellipses for petals and a small circle at the center of the petals (Figure 10a). This was based on Guideline 1, 4 and 5. P5 was able to somewhat associate the shape but enquired how the petals being connected to a plant. We quickly drew a curved line segment (Guideline 1 and 4) emerging from the center. To this P5 reacted (Figure 10b):

P5 – "Yes this is more like it. Okay. This is how a flower is connected. Yes, this looks like a flower because from the middle it's like this only. Yes, the petals are also okay, it's a flower."

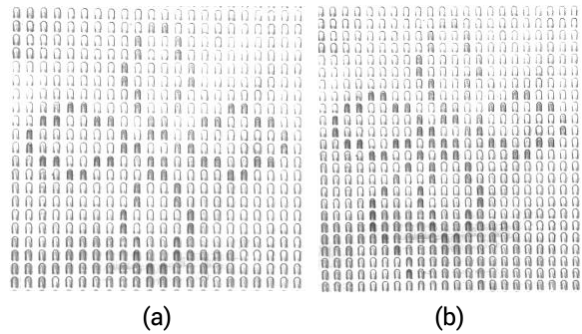

(a)                    (b)

Figure 10. (a) Initial flower, (b) Iterated flower

This demonstrates that Pixel Art guidelines are replicable to make any arbitrary shape but more importantly, the ability of the participants to imagine and suggest practical ways to iterate a tactile graphic, shows that participants can learn the interface and its capabilities.

### 5.4.2   *Learning through Pixelated Tactile Graphics*

The involvement of tactile readers with visual impairments in the process of designing tactile media through a fast, iterative pin-array display would improve the quality of tactile graphics [7]. The interaction also creates an opportunity for participants to learn, when they are supported by contextual frame of reference either from previous knowledge or new relatable information. We observed that participants could understand new concepts presented through Pixel Art on Tacilia. For example, P7 requested us to draw English characters. P7 had no previous knowledge about the shapes of English letters as all education for the participant had been in Braille or audio and hence, expressed an intent to learn alphabets on Tacilia. The researcher drew the initials of the university of the participant and the researchers "IIT and UCL" and then guided the participant through each letter by hand. P7 scanned and identified the shape and verbalized the characteristics of the pixelated line segments. The research informed P7 about the alphabet that was being felt, and through this procedure, P7 was able to learn the shapes of the English alphabet for the first time.

P7 – "This is a horizontal line above then a vertical line going down and then again, a horizontal line. What is this? (Researcher informs that this is an 'I'). Ah okay, 'I' is made like this and then this, this in another I."

As it can be seen in the above quote, P7 was able to recognize the letter 'I' at second instance without guidance. However, connecting the alphabet to create a word was not automatic and required further instruction to understand the gap between letters. It may be because reading among blind adults is limited, and their understanding of language is dependent on the phonetic assimilation of audio media rather than spellings and letters [33].

## 6   DISCUSSION

In this paper, we conducted a qualitative evaluation of the implementation of Pixel Art guidelines to create tactile shapes on a pin array display Tacilia and therefore our contribution takes this knowledge one step further as we identify how we should design these line segments and primitive shapes that would also combine to create more complex tactile graphics. The user study with eight blind and partially sighted tactile readers, in which, they compared the clarity of reading pixelated tactile shapes to raised line graphics revealed that it took greater time to read shapes on Tacilia, due to its low actuation height but shapes on both the mediums were read with comparable accuracy and in similar reading patterns. This shows that the use of the Pixel Art guidelines led to the design of intelligible basic tactile shapes on a pixelated display. We found that pixelated rectilinear and diagonal lines are clear to comprehend (Guideline 1, 2 and 3), corners are sharp (Guideline 6) and the curves, despite being difficult to make a pin array are also understandable (Guideline 5, 6). However, there were a few limitations to this implementation which are discussed later. Based on 64 video samples, we also report a four-step exploration procedure of reading tactile shapes. Finally, the refreshablity of Tacilia, combined with the guidance from Pixel Art created a new emerging interaction to design tactile graphics and learn from tactile media, which needs to be explored in future works. Overall, our findings have implications to improve accessibility through standardization, ease of implementation and interpretation that are discussed ahead.

## 6.1 Implications of Guidelines

Using guidelines to make tactile graphics can have several positive implications for the DIS community and towards accessibility of tactile information on pin arrays for people with visual impairments. Guidelines provide a set of recommendation to apply design principles to provide a positive user experience [60]. It provides a consistent framework for creating tactile shapes and establishes the conventions for representing lines and curves using a standardized approach. This standardization can also be evaluated for other pin array interfaces, which will eventually help users to develop a sense of familiarity with the representations of tactile graphics. Guidelines will also ensure that in the limited surface and resolution of refreshable pin array displays, image perception is consistent, accurate and fast. Various features (vertices, crosses, and proximal features) and line segment types (rectilinear, curved and diagonal) were explored in this research on a Tacilia prototype that participants were able to accurately identify with varying degrees of visual impairments. In larger, more advanced displays, the evaluation of the same guidelines is expected to provide a similar result, based on the Gestalt principle of good continuation [11] and the Law of Proximity [10].

Pixel Art guidelines for tactile displays were found to provide a necessary framework to design tactile graphics. The rules to create graphics enabled creativity, to express arbitrary tactile shapes that are comprehensible, leading to new tactile experiences and interactions. This translation of an arbitrary mental image into tactile shape following the given rules of Pixel Art can also be used as a framework of down-sampling algorithms for pin array displays. In addition, people can also be trained to create tactile graphics on a pin array display. By having a structured approach to the creation and presentation of tactile media, educators and trainers can align their mental image to the pixel grid for communicating information about shapes.

Building on the Frame of Interdependence for AT proposed by Bennett et al. [4], drawing and learning on Tacilia facilitated interdependence. The study showed scenarios in which the researcher learnt about *creating* tactile graphics. In contrast, the participants learnt about the *creations*. For example, P5's requirements were easily drawn and evaluated on the spot (Section 5.4.1). In another case, P7 got exposure to the characters of the English language for the first time using the pixelated display and was able to learn the shapes of few English alphabets. At the same time, P7 was critical of the juxtaposition of characters for instance and wished for more spacing in-between letters. This became new knowledge and presents a new question for the researchers, which will be explored in future works. Involvement of designers with visual impairments at the time of the creation of tactile graphics mediated by a reconfigurable pin-array display enables quick changes initiated by designers with visual impairments, which has been shown to improve the overall quality of tactile graphics [7]. Extending this phenomenon, by using Pixel Art, a consistent and replicable format of tactile graphics can be made for faster and easier interpretation. Hence, with the combination of refreshable pin arrays and Pixel Art designers of tactile media and its readers work together to codesign, iterate and learn from tactile media.

## 6.2 Beyond Pixel Art

In the scope of this paper, we presented qualitative evidence about the positive implementation of Pixel Art for basic line segments and shapes. These are the building blocks of any graphic or diagram. However, tactile graphics in everyday use feature many different line widths, textures, braille labels, tactile icons and symbols [51]. Graphics and diagrams can also compose of a combination of multiple basic shapes that are proximal to one another or cross in various orientations. These requirements may break some of the Pixel Art guidelines or would need more rules to present comprehensible information.

For example, P2 read the sine curve comfortably perhaps due to the familiarity, anticipation, and contextual sematic information about the diagram. However, for novice readers, a cluster of actuated tactile pixels at the origin of the graph may be disorientating. On another note, the semantic mapping of the position of English characters for P7 also took some further instructions, which was not part of the guidelines. The guidelines are hence limited to closed, simple shapes. For more complex diagrams where multiple features must be represented, or there is a need to create adjacent or crossing lines and textures, there is a need to go beyond the implementation of Pixel Art and generate new rules based on psychophysical studies. For instance, what should be the minimum gap between two distinct shapes to keep them distinct? How should we represent crossing lines or overlapping lines? How should we represent adjacent lines? Where should we put braille labels? and how to create textured areas? are some unanswered questions.

We also acknowledge that insights included in this paper are based on a limited sample of tactile graphics on a single pin-array type display. There were two reasons to limit our exploration at this stage. First, the Tacilia prototype only had 27x27 tactile pixels and therefore, only basic shapes with few details can be effectively presented in this small space of about 67.5 x 67.5mm. Secondly, we know that reading performance depends on previous exposure and experience [34]. The familiarity with tactile sensations of shapes, mental mapping of information to reconstruct visual images and regular exposure that enhances tactile discrimination skills are some of the individual factors that will influence the comprehension, speed and reading performance of tactile information [29,37,50]. Participants in the study had varying experiences with tactile graphics. Therefore, for a balanced comparison, we decided to limit our graphics to basic and simple geometric shapes that everyone can recognize to some extent. However, we did find that participants with previous exposure to tactile media performed better with Pixel Art, which is in line with previous research [18,43]. Despite the study having basic shapes and was conducted with a limited sample of participants, none of the participants had any experience in using pin array displays. Therefore, as each of the eight participants could read Pixel Art tactile shapes to a large extent, we can confidently say that the findings open the possibilities for the development of standards to create tactile graphics on refreshable pin array displays.

We encourage the HCI community to replicate and enhance the study by using the guidelines to create diverse tactile graphics on different display devices. Repeatable, standardized experiments across devices that include psychophysical and qualitative evaluation Pixel Art tactile graphic's size, orientation, gaps between pins and shapes, braille labels and graphical complexity will provide the necessary evidence to establish this as a standard method of rendering tactile graphics across pin array displays. We will also continue our exploration in future work to address the unanswered questions.

## 7 CONCLUSION

In conclusion, our paper presented a qualitative evaluation of the implementation of the Pixel Art system to refreshable pin array tactile displays. We made basic tactile shapes and graphics using the guidelines which eight tactile readers with visual impairments with varying levels experience positively evaluated to be clear and comprehensible, despite the limitations of the display. We also observed that the reading procedure for a tactile shape on a pin array has four sequential steps and new interdependent interactions emerge due to the refreshability of the display and the ease of implementing the Pixel Art guidelines. While our results highlight a variety of details about the tactile interaction, they also suggest unexplored gaps and unanswered questions. We hope that our results and ideas are utilized by researchers and encourage the community to further explore this underrepresented domain, to promote the development, usability, and adoption of tactile interfaces by people with visual impairments.

## ACKNOWLEDGMENTS

We would like to thank all the participants for their time and feedback and Sonali Jain to assist us with the recruitment. We would also like to thank Raised Line Foundation to provide us with their books. This research was supported by UK FCDO AT2030 Life-changing Access to Assistive Technologies (GB-GOV-1-300815) and UCL-IIT Delhi Strategic Partner Fund.

## REFERENCES

1. J. Abascal. 2002. Human-computer interaction in assistive technology: from "Patchwork" to "Universal Design." In *IEEE International Conference on Systems, Man and Cybernetics*, 6 pp. vol.3-. https://doi.org/10.1109/ICSMC.2002.1176076
2. Sandra Bardot, Marcos Serrano, Bernard Oriola, and Christophe Jouffrais. 2017. Identifying how Visually Impaired People Explore Raised-line Diagrams to Improve the Design of Touch Interfaces. In *Proceedings of the 2017 CHI Conference on Human Factors in Computing Systems*. Association for Computing Machinery, New York, NY, USA, 550–555. Retrieved September 3, 2021 from https://doi.org/10.1145/3025453.3025582
3. Yacine Bellik and Celine Clavel. 2017. Geometrical Shapes Rendering on a Dot-Matrix Display. In *Intelligent Human Computer Interaction*, Patrick Horain, Catherine Achard and Malik Mallem (eds.). Springer International Publishing, Cham, 8–18. https://doi.org/10.1007/978-3-319-72038-8_2
4. Cynthia L. Bennett, Erin Brady, and Stacy M. Branham. 2018. Interdependence as a Frame for Assistive Technology Research and Design. In *Proceedings of the 20th International ACM SIGACCESS Conference on Computers and Accessibility*, 161–173. https://doi.org/10.1145/3234695.3236348
5. T. Bhatnagar, N. Marquardt, M. Miodownik, and C. Holloway. 2021. Transforming a Monolithic Sheet of Nitinol into a Passive Reconfigurable Tactile Pixel Array Display at Braille Resolution. *In: 2021 IEEE World Haptics Conference (WHC). (pp. pp. 409-414). IEEE (2021) (In press).*, 409–414. Retrieved August 5, 2021 from https://ieeexplore.ieee.org/xpl/conhome/1001635/all-proceedings
6. T. Bhatnagar, V. Upadhyay, A. Sharma, P. V. M. Rao, M. Miodownik, N. Marquardt, and C. Holloway. 2021. Drawing Erasable Tactile Diagrams on Tacilia. *In: 2021 IEEE World Haptics Conference (WHC). (pp. pp. 593-594). IEEE (2021) (In press).*, 593–594. Retrieved August 5, 2021 from https://ieeexplore.ieee.org/xpl/conhome/1001635/all-proceedings
7. Jens Bornschein, Denise Prescher, and Gerhard Weber. 2015. Collaborative Creation of Digital Tactile Graphics. In *Proceedings of the 17th International ACM SIGACCESS Conference on Computers and Accessibility* (ASSETS '15), 117–126. https://doi.org/10.1145/2700648.2809869
8. J. E. Bresenham. 1965. Algorithm for computer control of a digital plotter. *IBM Systems Journal* 4, 1: 25–30. https://doi.org/10.1147/sj.41.0025
9. Matthew J. Burton, Jacqueline Ramke, Ana Patricia Marques, Rupert R. A. Bourne, Nathan Congdon, Iain Jones, Brandon A. M. Ah Tong, Simon Arunga, Damodar Bachani, Covadonga Bascaran, Andrew Bastawrous, Karl Blanchet, Tasanee Braithwaite, John C. Buchan, John Cairns, Anasaini Cama, Margarida Chagunda, Chimgee Chuluunkhuu, Andrew Cooper, Jessica Crofts-Lawrence, William H. Dean, Alastair K. Denniston, Joshua R. Ehrlich, Paul M. Emerson, Jennifer R. Evans, Kevin D. Frick, David S. Friedman, João M. Furtado, Michael M. Gichangi, Stephen Gichuhi, Suzanne S. Gilbert, Reeta Gurung, Esmael Habtamu, Peter Holland, Jost B. Jonas, Pearse A. Keane, Lisa Keay, Rohit C. Khanna, Peng Tee Khaw, Hannah Kuper, Fatima Kyari, Van C. Lansingh, Islay Mactaggart, Milka M. Mafwiri, Wanjiku Mathenge, Ian McCormick, Priya Morjaria, Lizette Mowatt, Debbie Muirhead, Gudlavalleti V. S. Murthy, Nyawira Mwangi, Daksha B. Patel, Tunde Peto, Babar M. Qureshi, Solange R. Salomão, Virginia Sarah, Bernadetha R. Shilio, Anthony W. Solomon, Bonnielin K. Swenor, Hugh R. Taylor, Ningli Wang, Aubrey Webson, Sheila K. West, Tien Yin Wong, Richard Wormald, Sumrana Yasmin, Mayinuer Yusufu, Juan Carlos Silva, Serge Resnikoff, Thulasiraj Ravilla, Clare E. Gilbert,

Allen Foster, and Hannah B. Faal. 2021. The Lancet Global Health Commission on Global Eye Health: vision beyond 2020. *The Lancet Global Health* 9, 4: e489–e551. https://doi.org/10.1016/S2214-109X(20)30488-5

10. Dempsey Chang, Keith V. Nesbitt, and Kevin Wilkins. 2007. The gestalt principles of similarity and proximity apply to both the haptic and visual grouping of elements. In *Proceedings of the eight Australasian conference on User interface - Volume 64* (AUIC '07), 79–86.

11. Dempsey Chang, Keith V. Nesbitt, and Kevin Wilkins. 2007. The Gestalt Principle of Continuation Applies to both the Haptic and Visual Grouping of Elements. In *Second Joint EuroHaptics Conference and Symposium on Haptic Interfaces for Virtual Environment and Teleoperator Systems (WHC'07)*, 15–20. https://doi.org/10.1109/WHC.2007.113

12. Curtis Chong. 2017. The Orbit Reader 20: The Most Inexpensive Braille Display. *Braille Monitor* 6, 9.

13. James D. Foley, Foley Dan Van, Andries Van Dam, Steven K. Feiner, and John F. Hughes. 1996. *Computer Graphics: Principles and Practice*. Addison-Wesley Professional.

14. Timothy Gerstner, Doug DeCarlo, Marc Alexa, Adam Finkelstein, Yotam Gingold, and Andrew Nealen. 2013. Pixelated image abstraction with integrated user constraints. *Computers & Graphics* 37, 5: 333–347. https://doi.org/10.1016/j.cag.2012.12.007

15. Clare Gilbert and Allen Foster. 2001. Childhood blindness in the context of VISION 2020 — The Right to Sight. *Bulletin of the World Health Organization*: 6.

16. Jiangtao Gong, Wenyuan Yu, Long Ni, Yang Jiao, Ye Liu, Xiaolan Fu, and Yingqing Xu. 2020. ”I can’t name it, but I can perceive it” Conceptual and Operational Design of ”Tactile Accuracy” Assisting Tactile Image Cognition. In *The 22nd International ACM SIGACCESS Conference on Computers and Accessibility* (ASSETS '20), 1–12. https://doi.org/10.1145/3373625.3417015

17. HanChu, WenQiang, HeShengfeng, ZhuQianshu, TanYinjie, HanGuoqiang, and WongTien-Tsin. 2018. Deep unsupervised pixelization. *ACM Transactions on Graphics (TOG)*. https://doi.org/10.1145/3272127.3275082

18. Morton A. Heller. 2016. Picture and Pattern Perception in the Sighted and the Blind: The Advantage of the Late Blind. *Perception*. https://doi.org/10.1068/p180379

19. Catherine Holloway, Dafne Zuleima Morgado Ramirez, Tigmanshu Bhatnagar, Ben Oldfrey, Priya Morjaria, Soikat Ghosh Moulic, Ikenna D. Ebuenyi, Giulia Barbareschi, Fiona Meeks, Jessica Massie, Felipe Ramos-Barajas, Joanne McVeigh, Kyle Keane, George Torrens, P. V.M. Rao, Malcolm MacLachlan, Victoria Austin, Rainer Kattel, Cheryl D Metcalf, and Srinivasan Sujatha. 2021. A review of innovation strategies and processes to improve access to AT: Looking ahead to open innovation ecosystems. *Assistive Technology* 33, sup1: 68–86. https://doi.org/10.1080/10400435.2021.1970653

20. Tiffany C. Inglis and Craig S. Kaplan. 2012. Pixelating vector line art. In *ACM SIGGRAPH 2012 Posters on - SIGGRAPH '12*, 1. https://doi.org/10.1145/2342896.2343021

21. Tiffany C. Inglis, Daniel Vogel, and Craig S. Kaplan. 2013. Rasterizing and antialiasing vector line art in the pixel art style. In *Proceedings of the Symposium on Non-Photorealistic Animation and Rendering* (NPAR '13), 25–32. https://doi.org/10.1145/2486042.2486044

22. Jeffrey James. 2013. The diffusion of IT in the historical context of innovations from developed countries. *Social indicators research* 111, 1: 175–184.

23. Sam Keddy. 2016. Pixel Art Outlines Tutorial. *Sam Keddy*. Retrieved August 15, 2021 from http://samkeddy.com/pixel-art-outlines/

24. R.L. Klatzky and S.J. Lederman. 1987. The intelligent hand. *The psychology of learning and motivation* 21: 121–151.

25. Ming-Hsun Kuo, Yong-Liang Yang, and Hung-Kuo Chu. 2016. Feature-Aware Pixel Art Animation. *Computer Graphics Forum* 35, 7: 411–420.

26. Susan J. Lederman, Roberta L. Klatzky, Cynthia Chataway, and Craig D. Summers. 1990. Visual mediation and the haptic recognition of two-dimensional pictures of common objects. *Perception & Psychophysics* 47, 1: 54–64. https://doi.org/10.3758/BF03208164

27. Gordon E. Legge, Cindee Madison, Brenna N. Vaughn, Allen M.Y. Cheong, and Joseph C. Miller. 2008. RETENTION OF HIGH TACTILE ACUITY THROUGHOUT THE LIFESPAN IN BLINDNESS. *Perception & psychophysics* 70, 8: 1471–1488. https://doi.org/10.3758/PP.70.8.1471

28. makegames. Pixel Art Tutorial. *Make Games*. Retrieved June 29, 2022 from https://makegames.tumblr.com/post/42648699708/pixel-art-tutorial

29. William Mccarthy and Joan Oliver. 1965. Some Tactile-Kinesthetic Procedures for Teaching Reading to Slow Learning Children. *Exceptional Children* 31, 8: 419–421. https://doi.org/10.1177/001440296503100805

30. Sile O'Modhrain, Nicholas Giudice, John Gardner, and Gordon Legge. 2015. Designing Media for Visually-Impaired Users of Refreshable Touch Displays: Possibilities and Pitfalls. *IEEE transactions on haptics* 8. https://doi.org/10.1109/TOH.2015.2466231

31. Krista E. Overvliet and Ralf Th. Krampe. 2018. Haptic two-dimensional shape identification in children, adolescents, and young adults. *Journal of Experimental Child Psychology* 166: 567–580. https://doi.org/10.1016/j.jecp.2017.09.024

32. Joyojeet Pal, Priyank Chandra, Terence O'Neill, Maura Youngman, Jasmine Jones, Ji Hye Song, William Strayer, and Ludmila Ferrari. 2016. An Accessibility Infrastructure for the Global South. In *Proceedings of the Eighth International Conference on Information and Communication Technologies and Development* (ICTD '16), 1–11. https://doi.org/10.1145/2909609.2909666

33. Konstantinos S. Papadopoulos, Evmorfia K. Arvaniti, Despina I. Dimitriadi, Vasiliki G. Gkoutsioudi, and Christina I. Zantali. 2009. Spelling performance of visually impaired adults. *British Journal of Visual Impairment* 27, 1: 49–64. https://doi.org/10.1177/0264619608097746

34. Delphine Picard and Samuel Lebaz. 2012. Identifying Raised-Line Drawings by Touch: A Hard but Not Impossible Task. *Journal of Visual Impairment & Blindness* 106, 7: 427–431. https://doi.org/10.1177/0145482X1210600705

35. Denise Prescher, Jens Bornschein, Wiebke Köhlmann, and Gerhard Weber. 2018. Touching graphical applications: bimanual tactile interaction on the HyperBraille pin-matrix display. *Universal Access in the Information Society* 17, 2: 391–409. https://doi.org/10.1007/s10209-017-0538-8

36. Bernhard Schmitz and Thomas Ertl. Interactively Displaying Maps on a Tactile Graphics Display.

37. Anchal Sharma, P V Madhusudhan Rao, and Srinivasan Venkataraman. 2022. Object recognition from two-dimensional tactile graphics: What factors lead to successful identification through touch? In *Proceedings of the 24th International ACM SIGACCESS Conference on Computers and Accessibility* (ASSETS '22), 1–5. https://doi.org/10.1145/3517428.3550376

38. Daniel Silber. 2015. *Pixel Art for Game Developers*. CRC Press.

39. Roger O. Smith, Marcia J. Scherer, Rory Cooper, Diane Bell, David A. Hobbs, Cecilia Pettersson, Nicky Seymour, Johan Borg, Michelle J. Johnson, Joseph P. Lane, S. Sujatha, P. V. M. Rao, Qussai M. Obiedat, Malcolm MacLachlan, and Stephen Bauer. 2018. Assistive technology products: a position paper from the first global research, innovation, and education on assistive technology (GREAT) summit. *Disability and Rehabilitation: Assistive Technology* 13, 5: 473–485. https://doi.org/10.1080/17483107.2018.1473895

40. WATANABE Tetsuya and OOUCHI Susumu. A Study of Legible Braille Patterns on Capsule Paper: Diameters of Braille Dots and Their Interspaces on Original Ink-printed Paper. 10.

41.  Gregg C. Vanderheiden. 1993. Making software more accessible for people with disabilities: a white paper on the design of software application programs to increase their accessibility for people with disabilities. *ACM SIGCAPH Computers and the Physically Handicapped*, 47: 2–32. https://doi.org/10.1145/155824.155826

42.  R. Velazquez, E. Pissaloux, M. Hafez, and J. Szewczyk. 2005. A low-cost highly-portable tactile display based on shape memory alloy micro-actuators. In *IEEE Symposium on Virtual Environments, Human-Computer Interfaces and Measurement Systems, 2005.*, 6. https://doi.org/10.1109/VECIMS.2005.1567577

43.  Catherine Y. Wan, Amanda G. Wood, David C. Reutens, and Sarah J. Wilson. 2010. Congenital blindness leads to enhanced vibrotactile perception. *Neuropsychologia* 48, 2: 631–635. https://doi.org/10.1016/j.neuropsychologia.2009.10.001

44.  Maarten W A Wijntjes, Thijs van Lienen, Ilse M Verstijnen, and Astrid M L Kappers. 2008. The Influence of Picture Size on Recognition and Exploratory Behaviour in Raised-Line Drawings. *Perception* 37, 4: 602–614. https://doi.org/10.1068/p5714

45.  Jacob O. Wobbrock, Shaun K. Kane, Krzysztof Z. Gajos, Susumu Harada, and Jon Froehlich. 2011. Ability-Based Design: Concept, Principles and Examples. *ACM Transactions on Accessible Computing* 3, 3: 1–27. https://doi.org/10.1145/1952383.1952384

46.  World Wide Web Consortium (W3C). 2015. Standards - W3C. *W3C*. Retrieved January 4, 2016 from http://www.w3.org/standards/

47.  Kaixing Zhao, Sandra Bardot, Marcos Serrano, Mathieu Simonnet, Bernard Oriola, and Christophe Jouffrais. 2021. Tactile Fixations: A Behavioral Marker on How People with Visual Impairments Explore Raised-line Graphics. In *Proceedings of the 2021 CHI Conference on Human Factors in Computing Systems* (CHI '21), 1–12. https://doi.org/10.1145/3411764.3445578

48.  2014. *Teaching Tactile Graphics*. Retrieved February 9, 2022 from https://www.youtube.com/watch?v=ZREwnV_XRsA

49.  2020. *Best Practices for Pixel Art*. CRC Press. https://doi.org/10.1201/9780429274596-19

50.  How to make tactile pictures understandable to the blind reader. Retrieved July 27, 2021 from https://www.dinf.ne.jp/doc/english/Us_Eu/conf/z19/z19001/z1900116.html#main

51.  BANA Home Page. Retrieved May 6, 2020 from http://www.brailleauthority.org/

52.  Product Narrative: Digital Assistive Technology | AT2030 Programme. Retrieved May 16, 2021 from https://at2030.org/product-narrative:-digital-assistive-technology/

53.  *Disability Interactions*. Retrieved July 5, 2022 from https://link.springer.com/book/10.1007/978-3-031-03759-7

54.  Dot Pad — The first tactile graphics display for the visually impaired. *Dot Pad*. Retrieved July 5, 2022 from https://pad.dotincorp.com/

55.  Two-dimensional, touch-sensitive graphic displays - metec AG. Retrieved July 5, 2022 from https://metec-ag.de/en/produkte-graphik-display.php

56.  Standards. *UKAAF*. Retrieved April 18, 2020 from https://www.ukaaf.org/braille/standards/

57.  What are Gestalt Principles? *The Interaction Design Foundation*. Retrieved February 19, 2022 from https://www.interaction-design.org/literature/topics/gestalt-principles

58.  To What Extent Do Gestalt Grouping Principles Influence Tactile Perception? Retrieved July 28, 2021 from https://oce-ovid-com.libproxy.ucl.ac.uk/article/00006823-201107000-00002/HTML

59.  RLF| Home. Retrieved September 1, 2021 from https://raisedlines.org/

60.  What are Design Guidelines? *The Interaction Design Foundation*. Retrieved May 12, 2023 from https://www.interaction-design.org/literature/topics/design-guidelines